# Free-Breathing PDFF and R2* Quantification at 0.55T: Reduced Respiratory Motion Sensitivity and a Time-Resolved 4D MRI Approach

Jeffery Wong[1,2], Jingjia Chen[1,2], Ding Xia[3], Xiang Xu[3], Els Fieremans[1,2], Dmitry S. Novikov[1,2], Hersh Chandarana[1,2], Li Feng[1,2]

[1] Bernard and Irene Schwartz Center for Biomedical Imaging, Department of Radiology, New York University Grossman School of Medicine, New York, NY, USA.

[2] Center for Advanced Imaging Innovation and Research (CAI²R), Department of Radiology, New York University Grossman School of Medicine, New York, NY, USA.

[3] Biomedical Engineering and Imaging Institute (BMEII) and Department of Radiology, Icahn School of Medicine at Mount Sinai, New York, NY, United States

**Running Head: Time-Resolved Multi-Echo 4D MRI**

Address correspondence to:

Li Feng, PhD

Associate Professor

Center for Advanced Imaging Innovation and Research (CAI²R)

New York University Grossman School of Medicine

Email: Li.Feng@nyulangone.org

## Abstract

**Purpose:** Free-breathing multi-echo MRI enables hepatic PDFF/R2* quantification without breath-holds, but respiratory motion can substantially affect R2* estimation through motion-induced field variations. This study investigated whether respiratory motion sensitivity is reduced at 0.55T versus 3T and developed a time-resolved free-breathing 4D MRI approach for PDFF/R2* quantification without respiratory binning.

**Methods:** A navigator-embedded multi-echo stack-of-stars sequence was implemented. Ten subjects were scanned under free-breathing and deliberate bulk-motion conditions at 3T and 0.55T. Data were reconstructed using motion-averaged, motion-resolved, and time-resolved approaches. R2* and PDFF were compared across reconstructions and motion conditions. Quantitative performance was characterized in a PDFF/R2* phantom using Cartesian and radial multi-echo acquisitions.

**Results:** Phantom experiments showed close radial-Cartesian agreement for both R2* and PDFF quantification at 3T. At 0.55T, R2* showed good agreement over the low-to-moderate range but higher uncertainty for high R2* values, where larger radial-Cartesian differences were also observed for PDFF. R2* quantification was substantially less sensitive to respiratory motion at 0.55T than at 3T, whereas PDFF was relatively insensitive to respiratory motion at both field strengths. Under deliberate bulk motion, time-resolved reconstruction provided robust R2* and PDFF measurements, with particularly high consistency between free-breathing and bulk-motion acquisitions at 0.55T.

**Conclusion:** R2* quantification was less sensitive to respiratory motion at 0.55T than at 3T, while PDFF remained relatively insensitive to respiratory motion. Time-resolved reconstruction provided additional robustness to bulk motion for quantification of both parameters. Reduced susceptibility effects at 0.55T and time-resolved reconstruction provide complementary advantages for motion-robust free-breathing hepatic PDFF/R2* quantification.

## 1. Introduction

Quantitative MRI has become an important clinical tool for noninvasive assessment of chronic liver diseases[1,2]. Among different quantitative biomarkers, the apparent transverse relaxation rate (R2*) and proton-density fat fraction (PDFF) are two of the most established MRI biomarkers for hepatic iron overload and steatosis, respectively[3–17]. Importantly, R2* and PDFF can be quantified simultaneously from a single multi-echo gradient-echo acquisition. Their excellent reproducibility and quantitative accuracy have supported increasing adoption of these two imaging markers in both clinical practice and research[8,18–22].

Conventional PDFF/R2* quantification uses a single breath-hold multi-echo gradient-echo acquisition to avoid respiratory motion artifacts. While this approach provides accurate and reliable measurements in cooperative patients, it remains challenging for those unable to hold their breath, such as pediatric and elderly patients. To address this limitation, considerable efforts over the past decade have developed techniques for free-breathing PDFF/R2* quantification using radial multi-echo acquisitions[23–33]. Compared with conventional Cartesian sampling, radial sampling is inherently more robust to respiratory motion[34] and has enabled free-breathing liver MRI, particularly when combined with respiratory motion compensation techniques[25,28,29,31,35–38]. Interestingly, these studies have also revealed a major difference between the two quantitative biomarkers, showing that PDFF remains relatively insensitive to respiratory motion, whereas R2* exhibits substantially higher motion sensitivity[39]. This finding highlights respiratory motion as a major challenge for accurate R2* quantification during free breathing.

The higher motion sensitivity of R2* is primarily caused by respiration-induced magnetic field variations[39]. During respiration, changes in tissue-air interfaces affect the local susceptibility distribution, producing time-varying off-resonance fields that introduce additional intravoxel dephasing and thus artificially elevate measured R2* values. In contrast, PDFF is determined primarily by the relative amplitudes of water and fat signals and is therefore less affected by these transient off-resonance changes. Since susceptibility-induced off-resonance effects scale with magnetic field strength, imaging at lower field strengths has the potential to reduce respiratory motion-induced errors in free-breathing R2* quantification.

The recent introduction of whole-body 0.55T MRI systems[40,41] provides a unique opportunity to revisit PDFF/R2* quantification in the liver[42–44]. Reduced susceptibility effects at 0.55T are expected to mitigate motion-induced errors in R2* quantification. However, chemical-shift imaging at 0.55T also presents new challenges. The smaller water-fat frequency separation requires longer echo spacing and therefore longer acquisition times, making free-breathing PDFF/R2* imaging particularly beneficial at this field strength. In addition, the 80-cm wide-bore design of the 0.55T system used in this study improves patient comfort but may also provide higher freedom for bulk patient motion during these relatively long scans[45].

Existing free-breathing PDFF/R2* imaging methods primarily rely on respiratory motion compensation techniques, including self-gating and respiratory-resolved reconstruction[25,28,29,31,46]. Respiratory-resolved methods, in particular, sort the acquired data into different motion states using a respiratory signal[28,31,37]. However, nonperiodic motion, such as bulk patient movement, may not be adequately represented by the underlying motion signal, which can lead to suboptimal data sorting and image artifacts.

In this work, we developed a time-resolved free-breathing imaging approach for liver PDFF/R2* quantification at 0.55T that does not require explicit respiratory binning. Using this framework, we investigated how field strength affects the respiratory motion sensitivity of PDFF/R2* quantification by comparing measurements at 3T and 0.55T. We further assessed its performance under both free-breathing and bulk-motion conditions and compared it with motion-averaged and motion-resolved reconstructions.

## 2. Methods

### 2.1 Overall Study Design

This work developed a time-resolved free-breathing multi-echo acquisition and reconstruction framework for quantitative liver PDFF/R2* imaging and compared respiratory motion sensitivity at 3T and 0.55T. The framework was evaluated through three complementary studies. First, phantom experiments characterized the quantitative performance of PDFF/R2* at both field strengths. Second, volunteer experiments compared respiratory motion sensitivity between 3T and 0.55T. Finally, the framework was evaluated under normal free-breathing and deliberate bulk-motion conditions to assess robustness to nonperiodic motion.

All experiments were performed with institutional review board (IRB) approval, and written informed consent was obtained from all participants prior to imaging. 10 volunteers (4 male, 6 female; age range, 23-38 years) were prospectively recruited for the study. Each volunteer underwent imaging on both a 3T scanner (MAGNETOM Prisma, Siemens Healthineers) and a 0.55T scanner (MAGNETOM Free.Max, Siemens Healthineers). The average time between the scans at both field strengths was 56.2 ± 69 days.

## 2.2 Time-Resolved Free-Breathing Multi-Echo Acquisition and Reconstruction

### *2.2.1 Free-Breathing Radial Multi-Echo Acquisition with Cross-Navigators*

A modified spoiled gradient-echo (GRE) sequence with a stack-of-stars trajectory was implemented for free-breathing multi-echo imaging[24]. To provide reliable motion information during the acquisition, two dedicated cross-navigator (xNavi) spokes were incorporated at each acquisition angle, as shown in **Figure 1a**. Specifically, the final two spokes of each stack/rotation angle were consistently acquired at kz = 0 along the anterior-posterior (AP) and left-right (LR) directions, respectively, rather than following the conventional stack-of-stars sampling pattern. These two orthogonal xNavi spokes were used to characterize temporal motion throughout the acquisition and were excluded for image reconstruction.

Acquisition parameters were selected separately for each field strength. At 3T, six echoes were acquired at TE = [1.4, 3.0, 4.6, 6.2, 7.8, and 9.4] ms, with TR = 10.8 ms, field of view (FOV) = 380 × 380 × 144 $mm^3$, matrix size = 190 × 190 × 24, spatial resolution = 2.0 × 2.0 × 6.0 $mm^3$, flip angle = 10°. 1020 spokes were acquired in each imaging slice, and the total acquisition time was 3 min 18 s.

Because the water-fat frequency separation is smaller at 0.55T, water and fat accrue relative phase more slowly, requiring a longer echo spacing to achieve sufficient phase evolution for fat-water separation. Two echo-spacing settings were evaluated at 0.55T with six echoes. The shorter echo-spacing setting used TE = [1.96, 4.11, 6.26, 8.41, 10.6, 12.7] ms, TR = 14.6 ms, and acquisition time = 4 min 4 s. The longer echo-spacing setting used TE = [2.4, 6.4, 10.4, 14.4, 18.4, 22.4] ms, TR = 26.0 ms, and acquisition time = 8 min 38 s. Common imaging parameters for both settings included: FOV = 380 × 380 × 128 $mm^3$, matrix size = 190 × 190 × 16, spatial resolution of 2.0 × 2.0 × 8.0 $mm^3$, flip angle = 5°, and 1600 spokes per imaging slice. A larger slice thickness and an increased number of radial spokes were acquired to compensate for the lower SNR at 0.55T.

### *2.2.2 Time-Resolved Multi-Echo Reconstruction and Quantification*

The dynamic multi-echo image series was reconstructed using a low-rank temporal subspace approach with spatiotemporal total variation regularization[47–56]. With this method, the dynamic image series is represented in a low-dimensional subspace with reduced degrees of freedom for reconstruction.

Time-resolved reconstruction was performed on a slice-by-slice basis. For each slice, every two consecutive radial spokes were grouped into one temporal frame. This yielded a total of 510 dynamic frames for 3T imaging and 800 frames for 0.55T imaging. Based on the corresponding acquisition times, the temporal resolution was approximately 0.39s per 3D volume at 3T, 0.31s per 3D volume for the shorter echo-spacing setting at 0.55T, and 0.65s per 3D volume for the longer echo-spacing setting at 0.55T. The AP and LR navigators were concatenated for temporal basis estimation using principal component analysis (PCA).

Specifically, the dynamic image series $X$ was represented using a low-rank temporal subspace model:

$$X = U_K V_K$$

where $U_K$ represents the $K$ dominant temporal basis functions and $V_K$ represents the corresponding spatial coefficient maps[51,52]. Prior to iterative reconstruction, the acquired radial k-space data were shifted onto a Cartesian grid using self-calibrating GRAPPA operator gridding (GROG)[57,58]. The subsequent iterative reconstruction was then performed on the Cartesian grid. Coil sensitivity maps were estimated from the temporally averaged k-space data using the Walsh method[59]. The Unstreaking algorithm[60,61] was applied to suppress strong streaking artifacts in the reconstructed images.

Using the pre-estimated temporal basis, the spatial coefficient maps were reconstructed by solving:

$$\hat{V}_K = \arg\min_{V_K} \|E(U_K V_K) - d\|_2^2 + \lambda_s \|\nabla_s V_K\|_1 + \lambda_t \|\nabla_t (U_K V_K)\|_1$$

where $E$ denotes the multi-coil encoding operator incorporating coil sensitivities and Cartesian k-space sampling mask, $d$ represents the GROG-shifted Cartesian k-space data, and $\nabla_s$ and $\nabla_t$ denote the spatial and temporal finite-difference operators, respectively, with corresponding regularization parameters $\lambda_s$ and $\lambda_t$ selected empirically for each field strength. The final time-resolved image series was obtained as:

$$\hat{X} = U_K \hat{V}_K$$

Following reconstruction, a reference end-expiratory frame was first identified based on the position of the lung-liver boundary, where respiratory motion could be readily visualized. Using the time-resolved first-echo images, the structural similarity index (SSIM) between each temporal frame and the reference frame was calculated over the 3D volume. Frames with SSIM values above a predefined threshold were selected, and the corresponding frames from all six echoes were averaged to generate the final multi-echo image set for PDFF/R2* quantification (**Figure 1b**). Default SSIM thresholds of 0.98 and 0.95 were used at 3T and 0.55T, respectively. The lower threshold was used at 0.55T to account for the lower image SNR and retain sufficient frames for averaging. Frame selection was visually verified, and the SSIM threshold was relaxed, when necessary, within the ranges of 0.96-0.98 at 3T and 0.93-0.95 at 0.55T for data with bulk motion, to balance respiratory-state consistency and image quality. Water-fat separation, B0 field map estimation, and R2* mapping were then performed on the final multi-echo images using the Graphcut method[62] with a six-peak fat spectral model[63–65]. B0 and R2* maps were jointly estimated, and PDFF was calculated using the magnitude discrimination method[66].

### 2.3 Quantitative Performance at 3T and 0.55T

A commercially available quantitative MRI phantom (Calimetrix, Madison, WI, USA)[22,67] was scanned at both 3T and 0.55T to evaluate the quantitative performance of the proposed radial multi-echo acquisition. The phantom contains 16 vials spanning PDFF values of 0-30% and R2* values of 0-600 $s^{-1}$ measured at 3T. At each field strength, the phantom was imaged using both the proposed radial multi-echo acquisition and a clinical Cartesian multi-echo acquisition. For the phantom imaging at 0.55T, the echo times of the radial acquisition were slightly adjusted from those used in vivo to match those of the corresponding Cartesian acquisition. Specifically, the shorter echo-spacing setting for both radial and Cartesian acquisitions used TE = [2.8, 5.0, 6.2, 7.4, 9.6, 11.8] ms, and the longer echo-spacing setting used TE = [2.8, 6.8, 10.8, 14.8, 18.8, 22.8] ms.

PDFF and R2* maps were generated using the same water-fat separation and quantitative mapping approach described above. Quantitative measurements were obtained from each phantom vial, and agreement between the radial and Cartesian acquisitions was evaluated for both PDFF and R2*. Linear regression and concordance correlation coefficient (CCC) were computed to assess agreement between the two acquisition approaches. As R2* estimation may become less reliable for rapidly decaying signals with high R2* values, the Cramér-Rao lower bound (CRLB) was also calculated to characterize R2* estimation uncertainty across the range of R2* values and echo-spacing settings evaluated at 0.55T.

The two echo-spacing settings were compared in one subject at 0.55T to select the echo-spacing setting for all in vivo experiments. Specifically, B0, R2*, and PDFF maps obtained with the shorter and longer echo-spacing settings were visually compared to assess their effects on in vivo quantitative imaging.

### 2.4 Respiratory Motion Sensitivity at 3T and 0.55T

To evaluate the effects of respiratory motion on PDFF/R2* quantification and their dependence on magnetic field strength, all subjects underwent free-breathing acquisitions at both 3T and 0.55T. At each field strength, the acquired data were reconstructed using three approaches, including motion-averaged reconstruction, motion-resolved reconstruction, and the proposed time-resolved reconstruction. For the motion-averaged reconstruction, all acquired data were combined directly without motion compensation. Motion-resolved reconstruction followed the XD-GRASP approach[28,37]. Specifically, the acquired data were first sorted into four respiratory states, ranging from end-expiration to end-inspiration, based on a respiratory signal extracted from xNavi. Dynamic compressed sensing reconstruction was then performed on the sorted data to generate respiratory-resolved images, and the end-expiratory image was used for quantitative analysis. For the time-resolved reconstruction, end-expiratory frames were selected and averaged as described above. PDFF and R2* maps were generated from each reconstruction for subsequent comparison.

To further evaluate the dependence of PDFF/R2* quantification on the extent of respiratory motion, an additional analysis was performed using the time-resolved reconstructions from a representative subject at both field strengths. After iterative reconstruction, the SSIM threshold used for frame selection was progressively reduced, resulting in the inclusion of a broader range of respiratory positions in the averaged images. PDFF and R2* maps were generated at each threshold to evaluate the sensitivity of both quantitative measurements to increasing respiratory motion.

For cohort-level analysis, three liver ROIs were placed per subject, one on each of three representative axial slices, to measure PDFF and R2* for each reconstruction approach at both field strengths. Agreement across reconstruction approaches was assessed using Bland-Altman analysis, with each ROI treated as an individual measurement. For the comparison between 3T and 0.55T, measurements from the three ROIs were averaged per subject to evaluate differences in respiratory motion sensitivity.

### 2.5 Bulk-Motion Experiments

To evaluate the robustness of quantitative imaging in the presence of bulk motion, all volunteers underwent an additional free-breathing acquisition at both 3T and 0.55T. During each acquisition, volunteers were instructed to perform a deliberate lateral body shift about halfway through the scan while continuing to breathe normally. In this case, the acquired data contained both respiratory motion and bulk motion. The data were reconstructed using the same motion-averaged, motion-resolved, and time-resolved approaches as in the free-breathing condition, and PDFF/R2* maps were generated for each reconstruction.

The same ROI and Bland-Altman analyses were performed for the bulk-motion acquisitions to assess the effects of combined respiratory and bulk motion across the three reconstruction approaches at both field strengths. Bulk-motion measurements were also compared directly with those from the corresponding free-breathing acquisition, separately for the motion-resolved and time-resolved reconstructions.

# 3. Results

### 3.1 Quantitative Performance at 3T and 0.55T

**Figure 2** summarizes the quantitative performance of R2* and PDFF at 3T and 0.55T in the phantom experiments. At 3T, radial and Cartesian measurements showed excellent agreement for both R2* (CCC = 0.998, slope = 0.97) and PDFF (CCC = 0.982, slope = 0.99). At 0.55T, radial and Cartesian R2* measurements showed good agreement over the low-to-moderate R2* range for both echo-spacing settings, whereas greater differences and variability were observed for higher R2* values. The shorter echo-spacing setting showed a below-unity regression slope (CCC = 0.912, slope = 0.83), whereas the longer echo-spacing setting showed a slope closer to unity but greater variability for higher R2* values (CCC = 0.931, slope = 1.02). Consistent with these observations, CRLB analysis showed low R2* estimation uncertainty over the low-to-moderate range, with predicted standard deviations of about 3 $s^{-1}$ at R2* $\approx$ 50 $s^{-1}$ and 6–7 $s^{-1}$ at R2* $\approx$ 150 $s^{-1}$. Estimation uncertainty increased substantially for higher R2* values, particularly with the longer echo-spacing setting, which reached approximately 108 $s^{-1}$ for Cartesian acquisition and 81 $s^{-1}$ for radial acquisition at R2* $\approx$ 600 $s^{-1}$. The increased uncertainty for both acquisitions indicated that the reduced precision at high R2* was not specific to radial sampling. These high R2* values were substantially above the range observed in vivo at 0.55T. For PDFF, radial and Cartesian measurements also showed greater variability at 0.55T than at 3T. Agreement was relatively higher with the shorter echo-spacing setting (CCC = 0.918, slope = 0.98) than with the longer echo-spacing setting (CCC = 0.875, slope = 1.19), and larger differences and uncertainty were observed particularly in vials with higher R2* values. Additional context regarding the field-strength dependence of R2* in this phantom and its implications for the high-R2* results is provided in **Supporting Information Note**.

The two echo-spacing settings were further compared in a representative subject at 0.55T (**Supporting Information Figure S1**). The longer echo-spacing setting produced visually less noisy B0 field and R2* maps than the shorter echo-spacing setting, while PDFF maps were more comparable. Based on this observation, the longer echo-spacing setting was selected for subsequent in vivo experiments at 0.55T.

The effect of navigator selection on temporal subspace estimation was also evaluated (**Supporting Information Figure S2**). For R2*, temporal basis estimation using the concatenated AP and LR navigators provided visually improved image quality compared with using either navigator alone, with the differences being more apparent at 0.55T than at 3T. In contrast, PDFF maps showed smaller differences among the AP-only, LR-only, and combined AP+LR navigator reconstructions at both field strengths. Based on these observations, the concatenated AP and LR navigators were used to estimate the temporal subspace in all time-resolved reconstructions.

### 3.2 Respiratory Motion Sensitivity of R2* and PDFF at 3T and 0.55T

Representative R2* and PDFF maps under free-breathing conditions are shown in **Figure 3**. At 3T (left panel), the motion-averaged reconstruction showed visibly elevated R2* values compared with both motion-resolved and time-resolved reconstructions, while the latter two approaches provided visually comparable R2* maps. In contrast, substantially smaller differences among the three reconstruction approaches were observed for R2* maps at 0.55T (right panel). PDFF maps were visually consistent across all reconstructions at both field strengths, demonstrating substantially lower sensitivity of PDFF to respiratory motion.

The dependence of R2* and PDFF quantification on the extent of respiratory motion was further evaluated in a representative case shown in **Figure 4**. The SSIM threshold used for frame selection was progressively reduced, thereby including a broader range of respiratory positions in the averaged images. At 3T, R2* values progressively increased as larger respiratory motion was included. In contrast, substantially smaller changes in R2* were observed at 0.55T across the same range of motion contribution. PDFF showed substantially smaller changes than R2* at both field strengths.

Data analysis across all subjects further demonstrated the difference in respiratory motion sensitivity of R2* between 3T and 0.55T (**Figure 5**). At 3T (top row), motion-averaged R2* measurements showed a systematic positive bias relative to both time-resolved (bias 16.87 $s^{-1}$, 95% limits of agreement [LoA] −11.60 to 45.34 $s^{-1}$) and motion-resolved reconstructions (bias 21.26 $s^{-1}$, LoA −4.75 to 47.26 $s^{-1}$), whereas motion-resolved and time-resolved measurements showed much closer agreement (bias −4.38 $s^{-1}$, LoA −16.86 to 8.09 $s^{-1}$). At 0.55T (bottom row), the differences among all three reconstruction approaches were substantially smaller, with biases of 0.16 $s^{-1}$ (LoA −1.35 to 1.66 $s^{-1}$) for motion-averaged versus time-resolved, 0.51 $s^{-1}$ (LoA −1.04 to 2.05 $s^{-1}$) for motion-averaged versus motion-resolved, and −0.35 $s^{-1}$ (LoA −2.04 to 1.34 $s^{-1}$) for motion-resolved versus time-resolved.

The corresponding PDFF analysis across all subjects is shown in **Figure 6**. At 3T (top row), small reconstruction-dependent differences were observed, with a motion-averaged versus time-resolved bias of 0.43% (LoA −1.30 to 2.16%) and a motion-averaged versus motion-resolved bias of 0.85% (LoA −0.64 to 2.33%). These differences were further reduced at 0.55T (middle row), where the three reconstruction approaches showed close agreement, with a motion-averaged versus time-resolved bias of 0.05% (LoA −0.74 to 0.83%) and a motion-averaged versus motion-resolved bias of −0.14% (LoA −0.73 to 0.44%). PDFF was therefore relatively insensitive to respiratory motion, with even smaller reconstruction-dependent differences at 0.55T. PDFF measurements also showed good agreement between 3T and 0.55T across all three reconstruction approaches (bottom row), with cross-field biases of −0.34% (LoA −2.40 to 1.71%) for motion-averaged, 0.04% (LoA −1.22 to 1.30%) for time-resolved, and 0.64% (LoA −0.95 to 2.23%) for motion-resolved reconstruction.

### 3.3 Improved Robustness to Bulk Motion with Time-Resolved Reconstruction

**Figure 7** shows representative $R2^*$ and PDFF maps under deliberate bulk motion at 3T and 0.55T from the same subject shown under normal free-breathing conditions in **Figure 3**. In the presence of bulk motion, motion-averaged reconstruction showed substantial artifacts, while residual artifacts remained with motion-resolved reconstruction. Time-resolved reconstruction provided improved image quality and $R2^*$ and PDFF maps that were more consistent with those obtained under normal free-breathing conditions (**Figure 3**).

The impact of combined respiratory and bulk motion on $R2^*$ quantification across all subjects is shown in **Figure 8**. At both 3T and 0.55T, motion-averaged reconstruction yielded higher $R2^*$ values than time-resolved reconstruction, with a bias of 71.41 $s^{-1}$ (LoA

−44.26 to 187.08 $s^{-1}$) at 3T and 15.00 $s^{-1}$ (LoA −39.05 to 69.04 $s^{-1}$) at 0.55T. Motion-averaged R2* values were also higher than those obtained with motion-resolved reconstruction, with a bias of 57.39 $s^{-1}$ (LoA −43.14 to 157.93 $s^{-1}$) at 3T and 9.08 $s^{-1}$ (LoA −36.52 to 54.67 $s^{-1}$) at 0.55T. Compared with time-resolved reconstruction, motion-resolved reconstruction yielded higher R2* values, with a bias of 14.02 $s^{-1}$ (LoA −24.96 to 52.99 $s^{-1}$) at 3T and 5.92 $s^{-1}$ (LoA −6.46 to 18.30 $s^{-1}$) at 0.55T. Overall, the differences among reconstruction approaches were substantially larger at 3T than at 0.55T.

The corresponding PDFF analysis under deliberate bulk motion is shown in **Figure 9**. At both 3T and 0.55T, motion-averaged reconstruction showed substantial differences in PDFF relative to time-resolved reconstruction, with a bias of 15.74% (LoA −26.06 to 57.54%) at 3T and 7.57% (LoA −13.50 to 28.63%) at 0.55T. Motion-averaged reconstruction also showed substantial differences relative to motion-resolved reconstruction, with a bias of 15.51% (LoA −24.57 to 55.59%) at 3T and 2.50% (LoA −15.47 to 20.48%) at 0.55T. Motion-resolved and time-resolved PDFF measurements showed closer agreement, particularly at 3T, with a bias of 0.24% (LoA −6.88 to 7.36%) at 3T and 5.06% (LoA −16.67 to 26.79%) at 0.55T. In the cross-field comparisons, time-resolved reconstruction showed close agreement between 3T and 0.55T, with a bias of −0.06% (LoA −3.71 to 3.59%), whereas substantially larger differences were observed with the other two reconstructions. Although PDFF was relatively insensitive to respiratory motion under free breathing, deliberate bulk motion introduced substantial differences when motion was not accounted for.

The robustness of time-resolved and motion-resolved reconstructions was further evaluated by directly comparing R2* measurements between normal free-breathing and deliberate bulk-motion acquisitions, as shown in **Figure 10**. At 0.55T, time-resolved reconstruction showed close agreement between the two acquisitions, with a bias of 1.11 $s^{-1}$ (LoA −2.16 to 4.37 $s^{-1}$), compared with −5.16 $s^{-1}$ (LoA −17.61 to 7.28 $s^{-1}$) for motion-resolved reconstruction. At 3T, larger differences between free-breathing and bulk-motion measurements were observed with both approaches, with a bias of 6.98 $s^{-1}$ (LoA −24.30 to 38.26 $s^{-1}$) for time-resolved reconstruction and −11.42 $s^{-1}$ (LoA −37.03 to 14.19 $s^{-1}$) for motion-resolved reconstruction. Time-resolved R2* measurements were therefore highly consistent between the two motion conditions at 0.55T.

The corresponding comparison of PDFF between the free-breathing and bulk-motion acquisitions is shown in **Supporting Information Figure S3**. Time-resolved reconstruction showed close agreement between the two acquisitions at both field strengths, with a bias of −0.18% (LoA −4.02 to 3.66%) at 3T and −0.08% (LoA −1.96 to 1.79%) at 0.55T. In contrast, larger differences were observed with motion-resolved reconstruction, with a bias of −0.84% (LoA −6.87 to 5.20%) at 3T and −4.95% (LoA −26.92 to 17.01%) at 0.55T.

## Discussion

### 4.1 Reduced Motion Sensitivity of R2* at 0.55T

A major finding of this study is the substantially reduced sensitivity of R2* quantification to respiratory motion at 0.55T compared with 3T. Under free breathing, motion-averaged reconstruction at 3T yielded systematically higher R2* values than motion-resolved and time-resolved reconstructions, whereas differences among the three approaches were much smaller at 0.55T. The analysis in **Figure 4** further showed that R2* increased progressively as more respiratory motion was included at 3T, whereas the corresponding changes were much smaller at 0.55T. PDFF showed only small reconstruction-dependent differences at 3T, which were further reduced at 0.55T, indicating that respiratory motion predominantly affects R2*.

The different motion sensitivities of R2* and PDFF can be attributed to the distinct physical mechanisms underlying their estimation[39]. R2* reflects transverse signal decay and is sensitive to local magnetic field inhomogeneity. Respiratory motion changes the relative positions of the liver, diaphragm, and lungs, producing temporal variations in the local susceptibility distribution. When data from different respiratory positions are combined, these variations introduce additional dephasing and an apparent increase in R2*. In contrast, PDFF is derived primarily from the relative water and fat signal amplitudes and is less sensitive to motion-induced field variations. Nevertheless, respiratory motion can still introduce inconsistencies among the multi-echo signals used for water-fat separation. The smaller reconstruction-dependent PDFF differences observed at 0.55T may therefore also reflect reduced susceptibility-induced field variations at lower field strength.

Because susceptibility-induced frequency offsets scale approximately with B0, respiration-induced changes in the susceptibility distribution produce smaller local field variations at 0.55T than at 3T, and therefore less additional dephasing and a smaller apparent increase in R2*. This provides a physical advantage that complements motion-compensation techniques. Although advanced motion-compensation approaches can substantially reduce errors due to anatomic displacement and mixing of different respiratory positions at 3T, they may not fully account for dynamic susceptibility-induced field variations associated with respiration. At 0.55T, these field variations are intrinsically reduced, which may facilitate more robust free-breathing R2* quantification.

### 4.2 Improved Motion Robustness with Time-Resolved Reconstruction

The proposed time-resolved reconstruction provided additional robustness when respiratory motion was accompanied by nonperiodic bulk motion. Under normal free-breathing, both motion-resolved and time-resolved reconstructions provided comparable R2* measurements, indicating that respiratory-resolved reconstruction can effectively compensate for standard respiratory motion. However, in the presence of bulk motion, differences between the two approaches became more pronounced. Motion-averaged reconstruction yielded the highest R2* values at both field strengths, whereas motion-resolved reconstruction substantially reduced this overestimation yet still yielded slightly higher R2* values than the time-resolved approach. This pattern suggested that respiratory-resolved reconstruction effectively addresses periodic respiratory motion but may be less effective when additional nonperiodic motion is present.

This difference may reflect the ways these two approaches handle motion. Motion-resolved reconstruction relies on a respiratory signal to sort the acquired data into different respiratory states and assumes that the dominant motion is due to respiration. Nonperiodic bulk motion can disrupt this relationship, such that data acquired at different body positions may be assigned to the same respiratory state. In contrast, the time-resolved approach reconstructs a series of high-temporal-resolution 3D volumes without explicit respiratory binning, from which motion-consistent end-expiratory frames can be retrospectively selected and combined. This flexibility may explain the improved robustness observed in the bulk-motion experiments.

The bulk-motion experiments also provide additional insight into the motion sensitivity of PDFF. Although PDFF was relatively insensitive to respiratory motion under normal free breathing, larger differences among reconstruction approaches emerged with deliberate bulk motion, particularly between motion-averaged and the two motion-compensated reconstructions. Robustness to respiratory motion therefore does not imply insensitivity to body movement, which may be more likely at 0.55T given the wider bore. Nevertheless, motion-resolved and time-resolved reconstructions still showed substantially closer agreement than the motion-averaged reconstruction, demonstrating the importance of accounting for nonperiodic motion even for PDFF quantification.

Direct comparison between normal free-breathing and deliberate bulk-motion acquisitions (**Figure 10** and **Supporting Information Figure S3**) showed that time-resolved reconstruction maintained close agreement between the two motion conditions for R2* at 0.55T and for PDFF at both field strengths, whereas motion-resolved reconstruction showed larger differences. The particularly high consistency of R2* at 0.55T suggests complementary advantages of lower field strength and time-resolved reconstruction, which may provide a robust approach for free-breathing quantitative liver MRI in patients who cannot reliably hold their breath or remain in a fixed position during the exam.

**4.3 Quantitative Performance and Technical Tradeoffs at 0.55T**

Despite the reduced motion sensitivity at 0.55T, quantitative R2* and PDFF imaging at lower field strengths face several technical challenges. First, the reduced SNR at 0.55T can compromise image quality and quantitative precision. In this study, acquisition and reconstruction were designed to provide sufficient SNR, and further improvements may be possible with advanced reconstruction and denoising[41,44,68–71]. Second, the reduced water-fat frequency separation at 0.55T requires larger echo spacing to provide sufficient phase evolution for water-fat separation and field-map estimation. In vivo, the longer echo spacing yielded visually less noisy B0 and R2* maps and was therefore selected for the volunteer experiments. However, because R2* is intrinsically lower at 0.55T, signal decay across the echo train is correspondingly slower, which mitigates this concern at the lower field strength.

The phantom experiments further characterized this tradeoff. At 3T, radial and Cartesian measurements agreed closely for both R2* and PDFF across the evaluated range. At 0.55T, agreement was good over the low-to-moderate R2* range but degraded at high R2* for both acquisitions, consistent with the increased uncertainty predicted by the CRLB analysis and indicating that this behavior was not specific to radial sampling. PDFF variability at 0.55T was likewise greatest in vials with high R2*, likely because rapid signal decay reduces the usable signal at later echoes for water-fat separation and R2* quantification.

These findings should be considered in the context of the phantom R2* range, which was specified by the manufacturer up to 600 $s^{-1}$ at 3T, well into the range associated with substantial hepatic iron overload and substantially above the values measured in the healthy volunteers of this study. The increased estimation uncertainty at 0.55T was confined to these high-R2* vials, so the observed limitations may be less relevant for subjects without substantial hepatic iron overload. Nevertheless, because the relationship between hepatic iron concentration and R2* at 0.55T is less established than at 1.5T and 3T, evaluation in patients with hepatic iron overload will be important to determine the clinically relevant range over which R2* and PDFF can be robustly quantified at 0.55T. Additional information regarding the field-strength dependence of R2* in this phantom and its implications for the high-R2* results is provided in **Supporting Information Note**.

### 4.4 Limitations and Future Directions

This study has several limitations. First, the study included a relatively small cohort of volunteers, and the range of R2* and PDFF values was therefore limited. Additional evaluation in patients with hepatic steatosis and iron overload will be important to further establish the performance of the proposed approach across a broader range of disease severity.

Second, although the phantom experiments demonstrated good agreement between the radial and clinical Cartesian acquisitions, independent reference R2* values were not available at 0.55T. Therefore, the phantom experiments primarily establish consistency between the two acquisition approaches rather than absolute accuracy at 0.55T.

Third, this study focused on comparing 3T and 0.55T and did not include measurements at 1.5T. These two field strengths were selected to provide a clear comparison of respiratory motion sensitivity across substantially different susceptibility environments. Future studies across multiple field strengths could more comprehensively characterize this relationship.

Fourth, the free-breathing acquisition used in this study had a relatively long scan time, particularly at 0.55T with the longer echo-spacing setting. The acquisition was designed to provide sufficient data for quantification and to investigate motion sensitivity at the two field strengths, and the effect of reducing the number of radial spokes was not systematically evaluated. Future work is needed to shorten the acquisition while maintaining image quality and quantitative performance.

## 5. Conclusion

This study demonstrates that R2* quantification is substantially less sensitive to respiratory motion at 0.55T than at 3T, while PDFF remains relatively insensitive to respiratory motion at both field strengths. A time-resolved approach was also developed for free-breathing R2* and PDFF quantification, providing additional robustness to nonperiodic bulk motion, which may be particularly relevant for the wide-bore 0.55T system used in this study. Together, the reduced-susceptibility effects at lower field strengths and time-resolved reconstruction may provide complementary advantages for motion-robust free-breathing PDFF/R2* quantification.

## Acknowledgment

This work was supported in part by the NIH (R01EB030549, R01DK143170, and P41EB017183) and was performed under the rubric of the Center for Advanced Imaging Innovation and Research (CAI$^2$R), an NIBIB National Center for Biomedical Imaging and Bioengineering.

## Figures

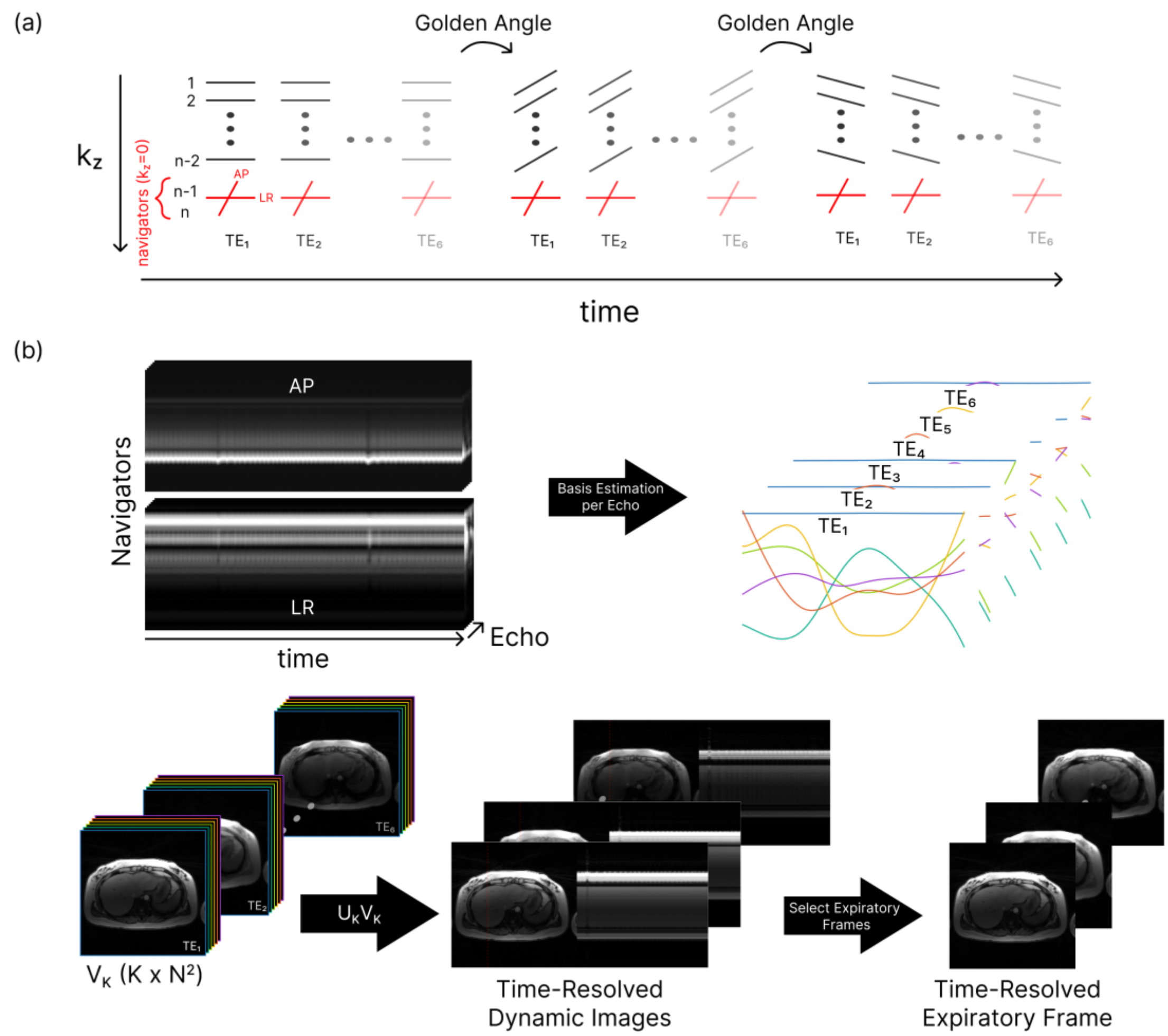

**Figure 1:** Overview of the time-resolved free-breathing multi-echo 4D MRI framework. (a) Modified multi-echo stack-of-stars sequence with cross-navigator (xNavi) spokes. At each acquisition angle, the final two spokes of each stack are acquired at kz = 0 along the anterior-posterior (AP) and left-right (LR) directions, respectively. (b) Time-resolved reconstruction and quantitative mapping framework. Temporal basis is estimated from the concatenated AP and LR xNavi signals and used for low-rank subspace reconstruction of a dynamic 3D multi-echo image series with sub-second temporal resolution. End-expiratory frames are identified using SSIM-based frame selection and averaged across the corresponding frames at all six echoes to generate the final multi-echo images for water-fat separation, B0 field map estimation, R2* mapping, and PDFF quantification.

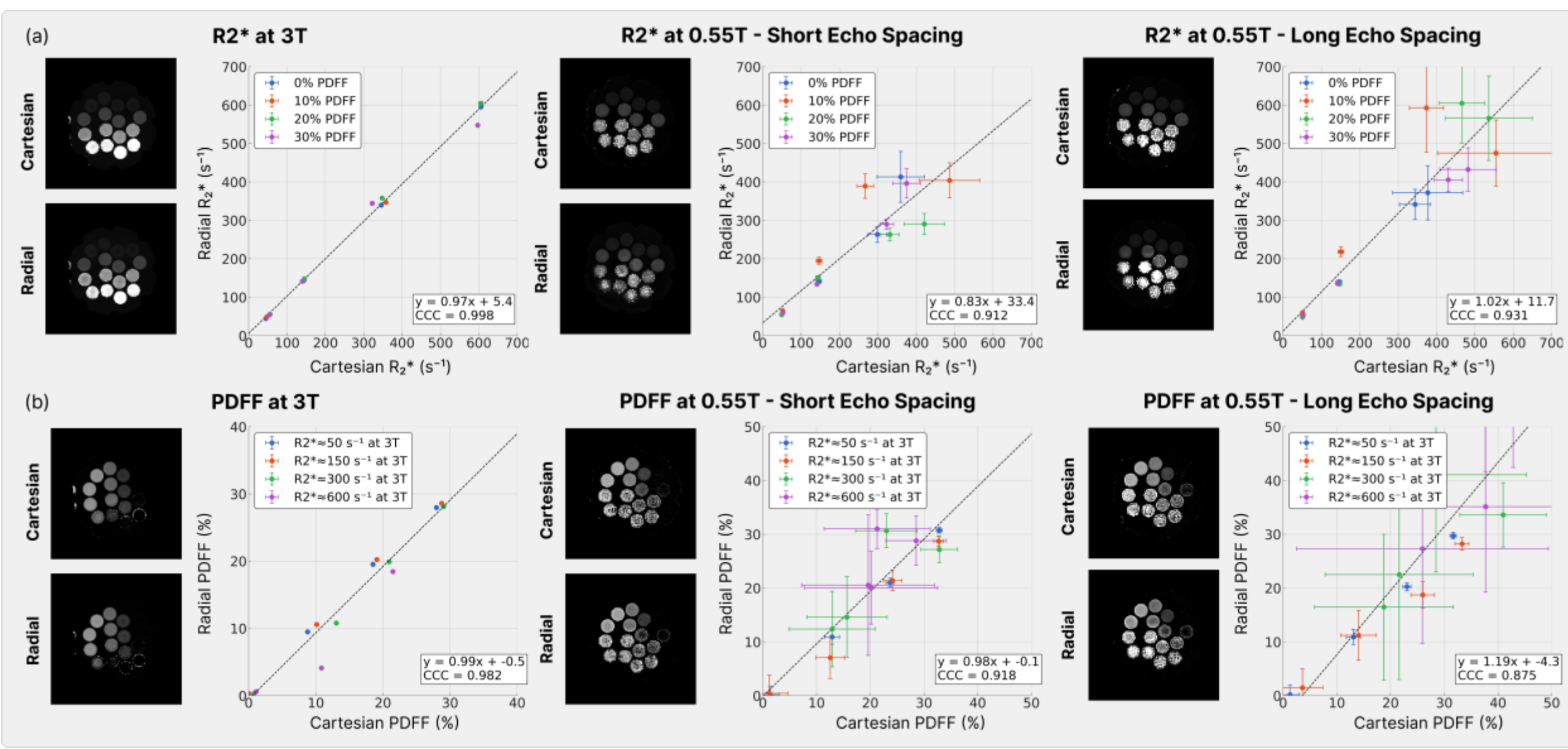

**Figure 2:** Phantom comparison of radial and Cartesian R2* and PDFF quantification at 3T and 0.55T. (a) R2* showed close radial-Cartesian agreement at 3T and good agreement over the low-to-moderate range at 0.55T, with larger differences for higher R2* values. CRLB analysis showed increased uncertainty in R2* estimates at higher R2* values, particularly with the longer echo-spacing setting. (b) PDFF showed close radial-Cartesian agreement at 3T, whereas larger differences and increased estimation uncertainty were observed at 0.55T, particularly in vials with higher R2* values. Error bars represent estimation uncertainty based on the CRLB.

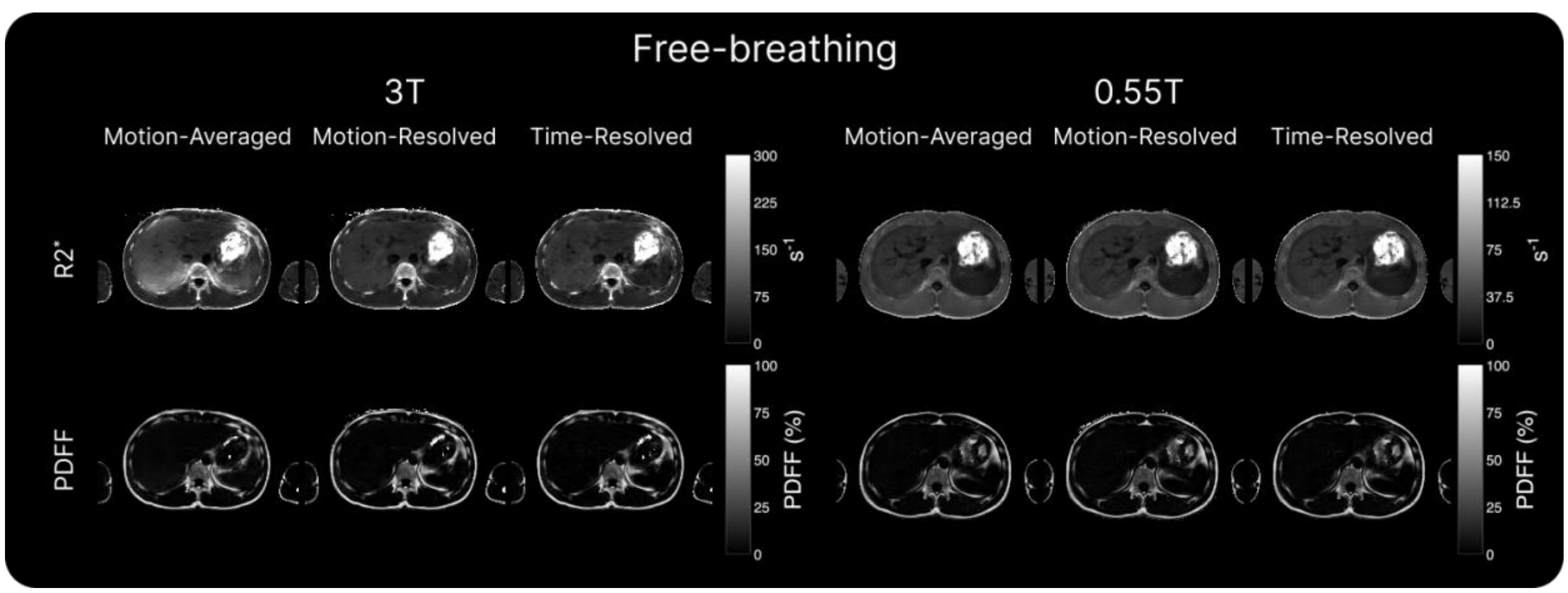

**Figure 3:** Representative R2* and PDFF maps under normal free-breathing conditions are shown for motion-averaged, motion-resolved, and time-resolved reconstructions at 3T (left) and 0.55T (right). At 3T, motion-averaged reconstruction shows visibly elevated R2* compared with motion-resolved and time-resolved reconstructions, whereas substantially smaller differences are observed at 0.55T. PDFF maps show smaller differences among the three reconstruction approaches at both field strengths.

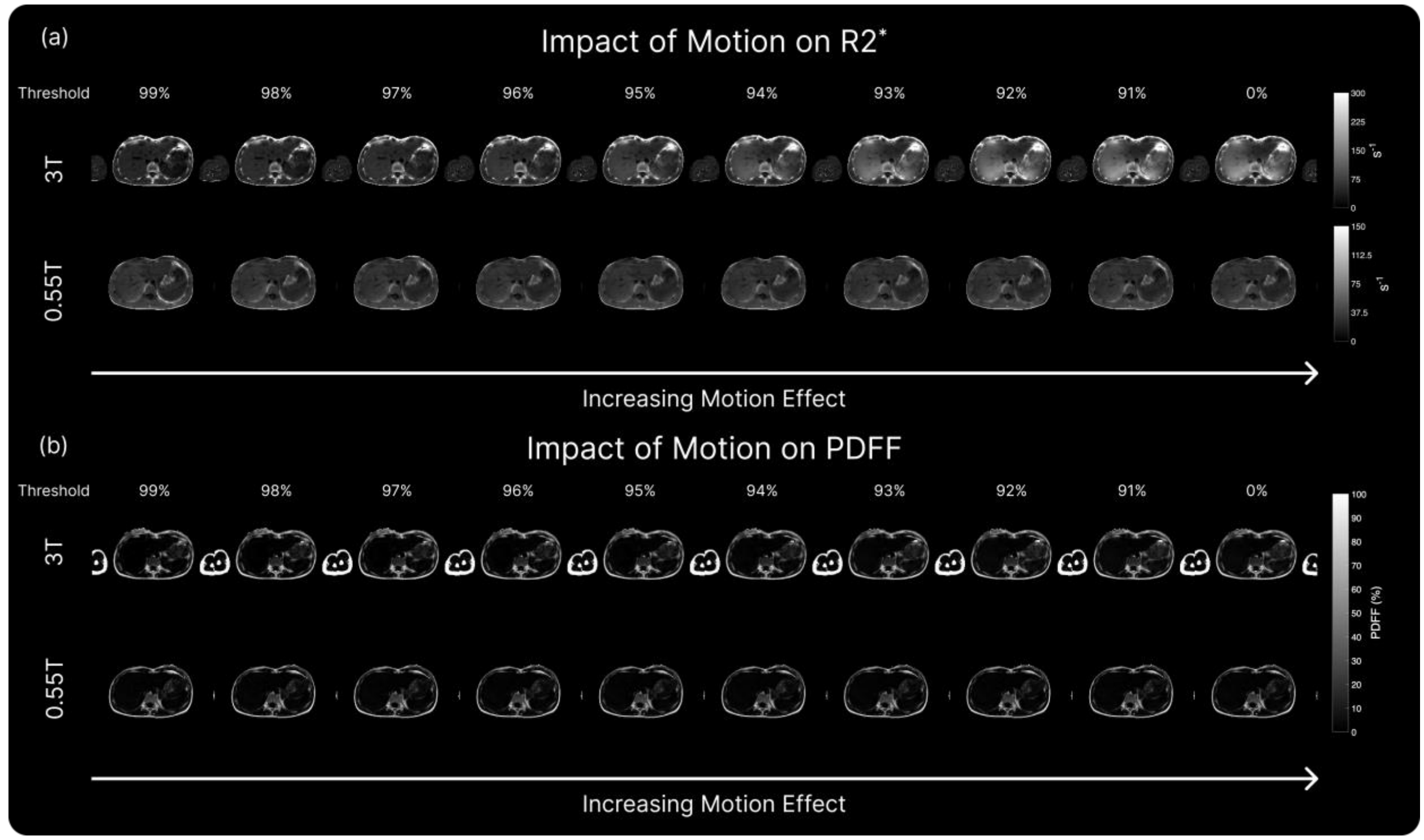

**Figure 4:** Representative R2* (a) and PDFF (b) maps are shown at 3T (top) and 0.55T (bottom). From left to right, progressively more temporal frames are included as the SSIM threshold for frame selection is lowered, resulting in greater contributions from different respiratory positions. At 3T, R2* progressively increases as a greater extent of respiratory motion is included, whereas substantially smaller changes are observed at 0.55T. PDFF shows substantially smaller changes than R2* at both field strengths.

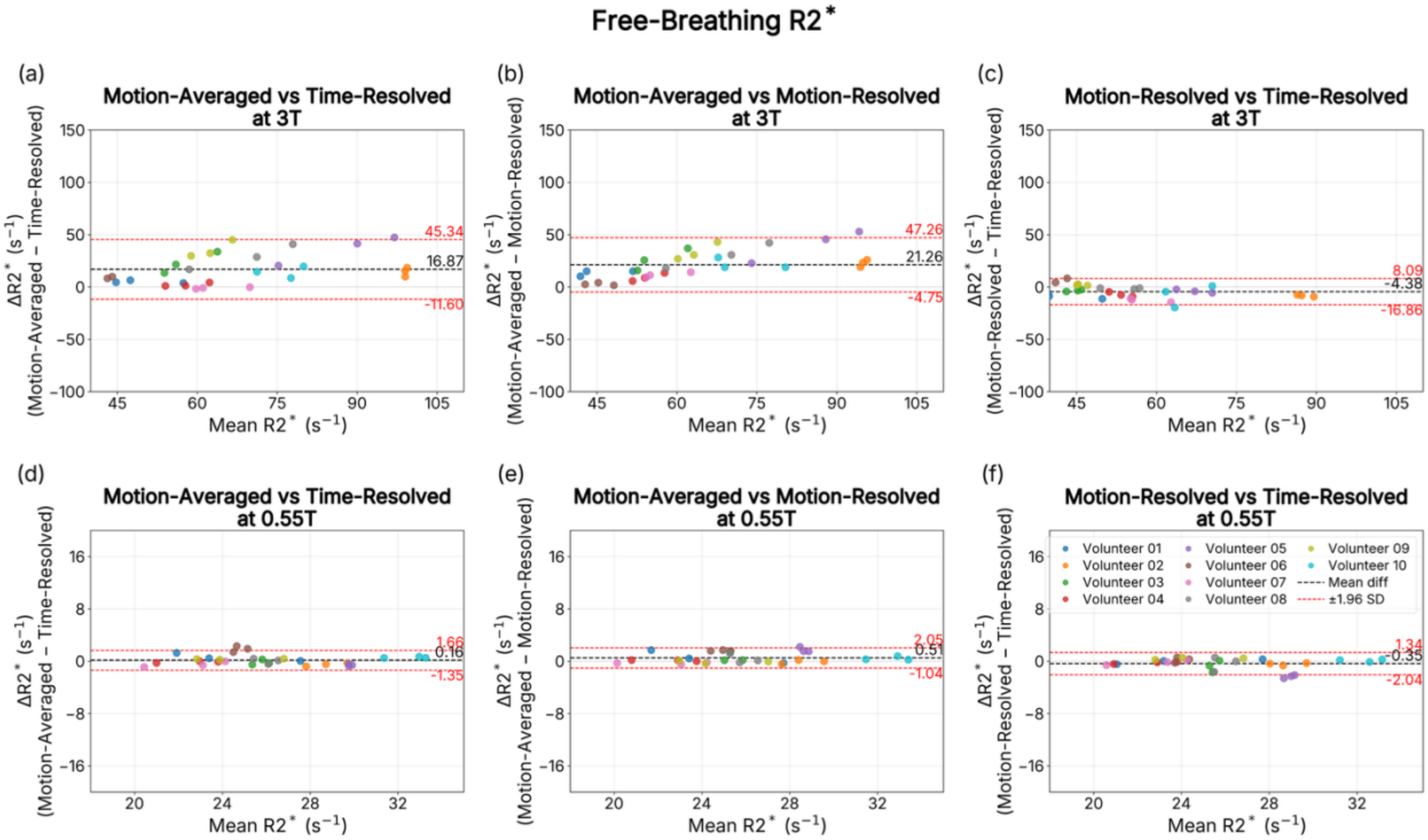

**Figure 5:** The top row shows comparisons among motion-averaged, motion-resolved, and time-resolved R2* measurements at 3T, and the bottom row shows the corresponding comparisons at 0.55T. Motion-averaged R2* measurements showed a systematic positive bias relative to motion-resolved and time-resolved measurements at 3T, whereas substantially smaller differences were observed at 0.55T.

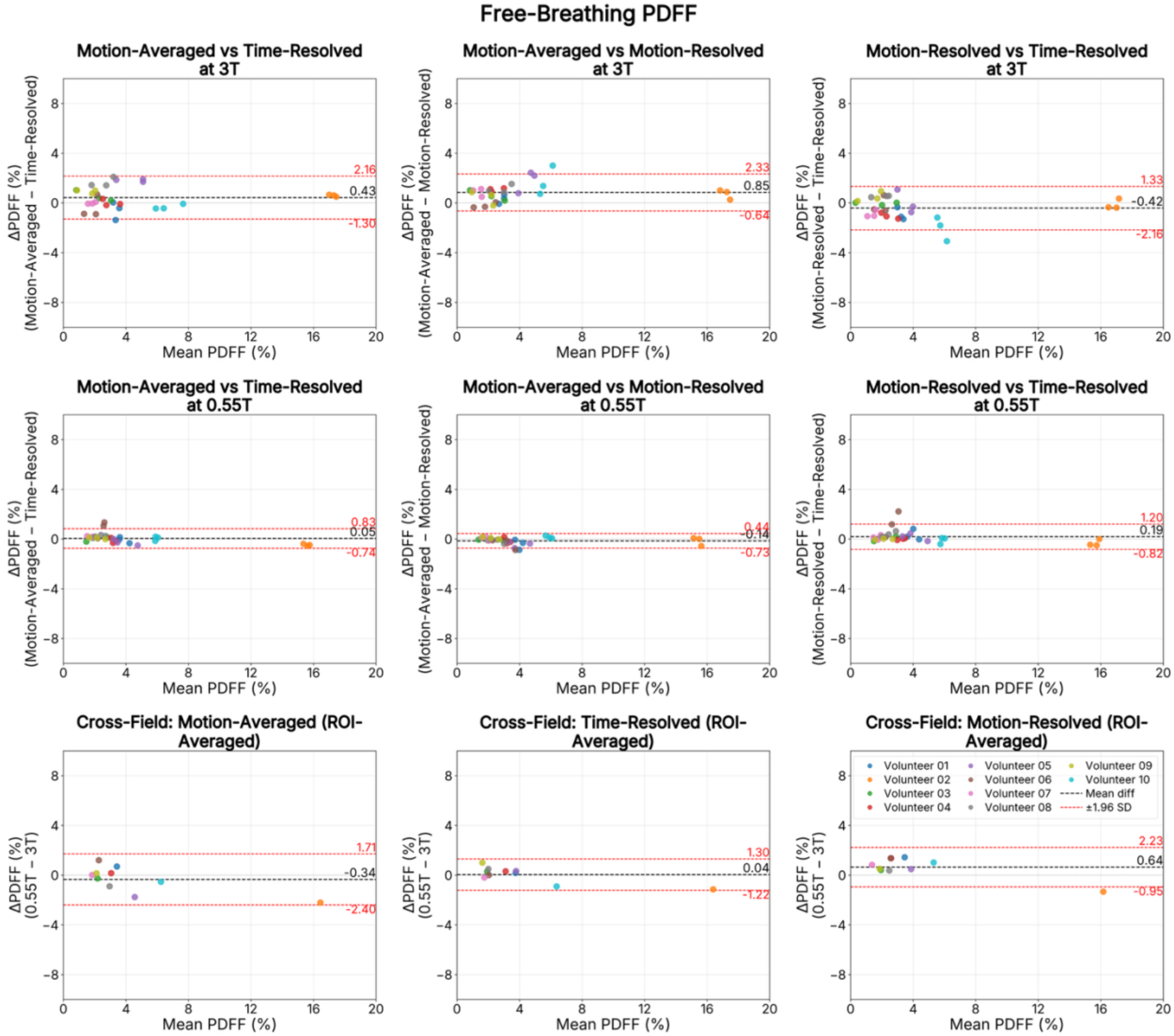

**Figure 6:** The top row shows comparisons among motion-averaged, motion-resolved, and time-resolved PDFF measurements at 3T, and the middle row shows the corresponding comparisons at 0.55T. Small reconstruction-dependent differences were observed at 3T, whereas these differences were further reduced at 0.55T. The bottom row compares PDFF measurements between 3T and 0.55T for the three reconstruction approaches.

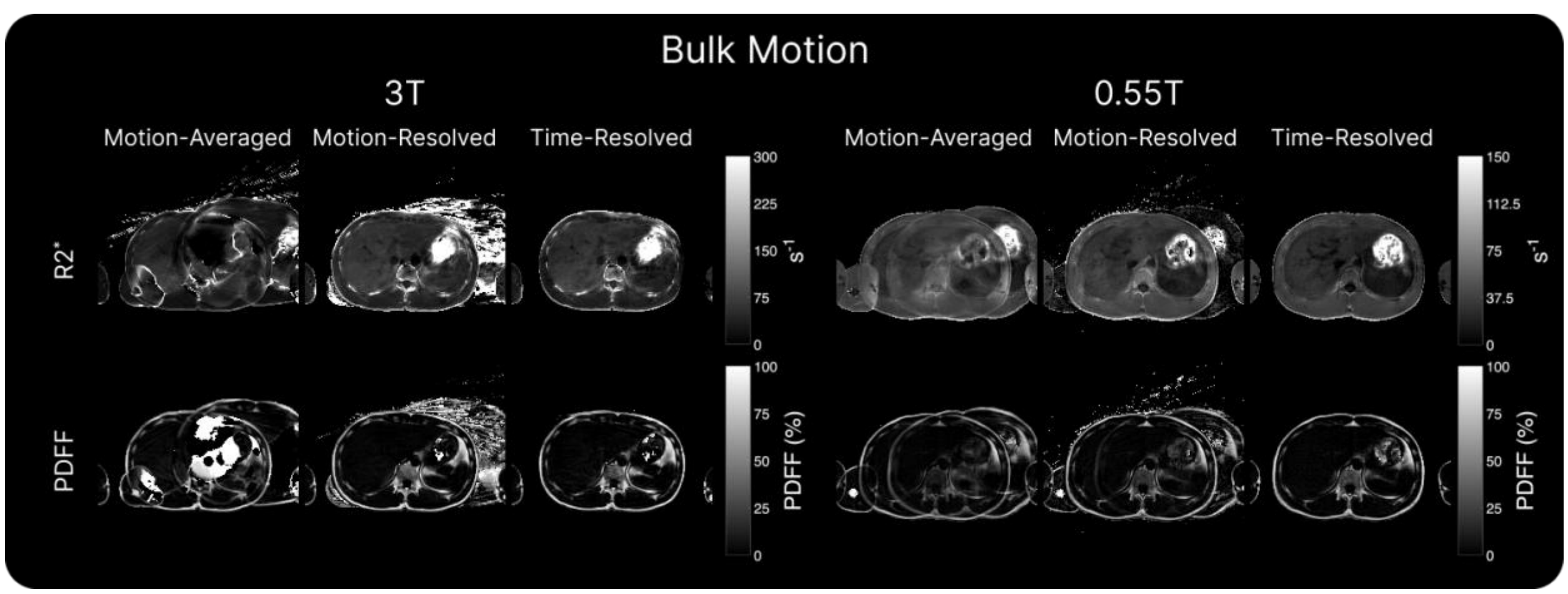

**Figure 7:** Representative R2* and PDFF maps under deliberate bulk-motion conditions are shown for motion-averaged, motion-resolved, and time-resolved reconstructions at 3T (left) and 0.55T (right). At both field strengths, motion-averaged reconstruction shows substantial artifacts, whereas residual artifacts persist in motion-resolved reconstruction. Time-resolved reconstruction provides improved image quality and R2* and PDFF maps that are more consistent with those obtained under normal free-breathing conditions, as shown in Figure 3.

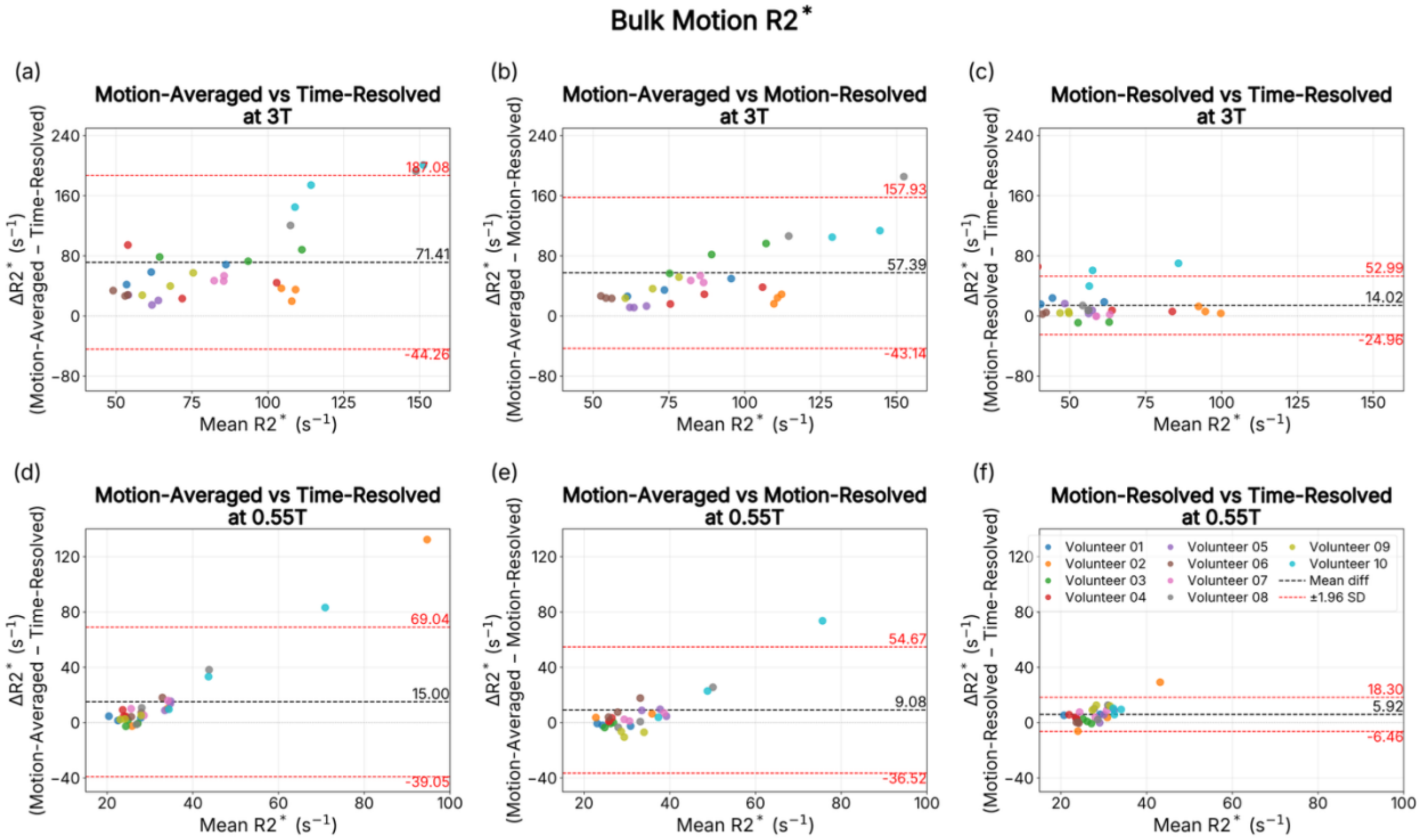

**Figure 8:** The top row shows comparisons among motion-averaged, motion-resolved, and time-resolved R2* measurements at 3T under deliberate bulk motion, and the bottom row shows the corresponding comparisons at 0.55T. At both field strengths, motion-averaged reconstruction yielded higher R2* values than time-resolved reconstruction and generally higher values than motion-resolved reconstruction. Motion-resolved reconstruction also yielded higher R2* values than time-resolved reconstruction, with larger differences observed at 3T than at 0.55T.

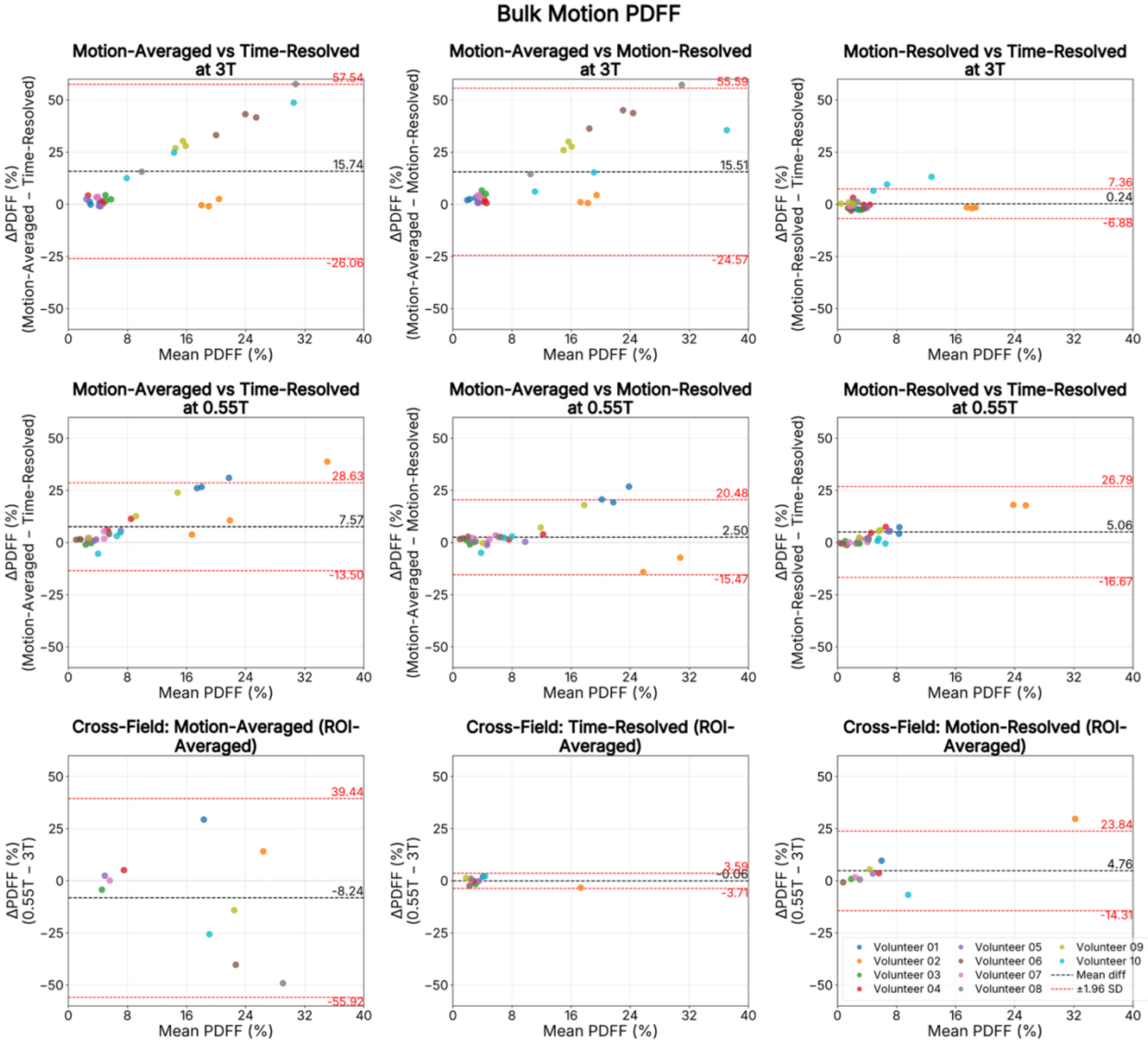

**Figure 9:** The top row shows comparisons among motion-averaged, motion-resolved, and time-resolved PDFF measurements at 3T under deliberate bulk motion, and the middle row shows the corresponding comparisons at 0.55T. At both field strengths, motion-averaged reconstruction showed a positive PDFF bias relative to motion-resolved and time-resolved reconstructions, with larger differences observed in subjects with higher PDFF. The bottom row compares PDFF measurements between 3T and 0.55T for the three reconstruction approaches. Time-resolved reconstruction showed the closest cross-field agreement.

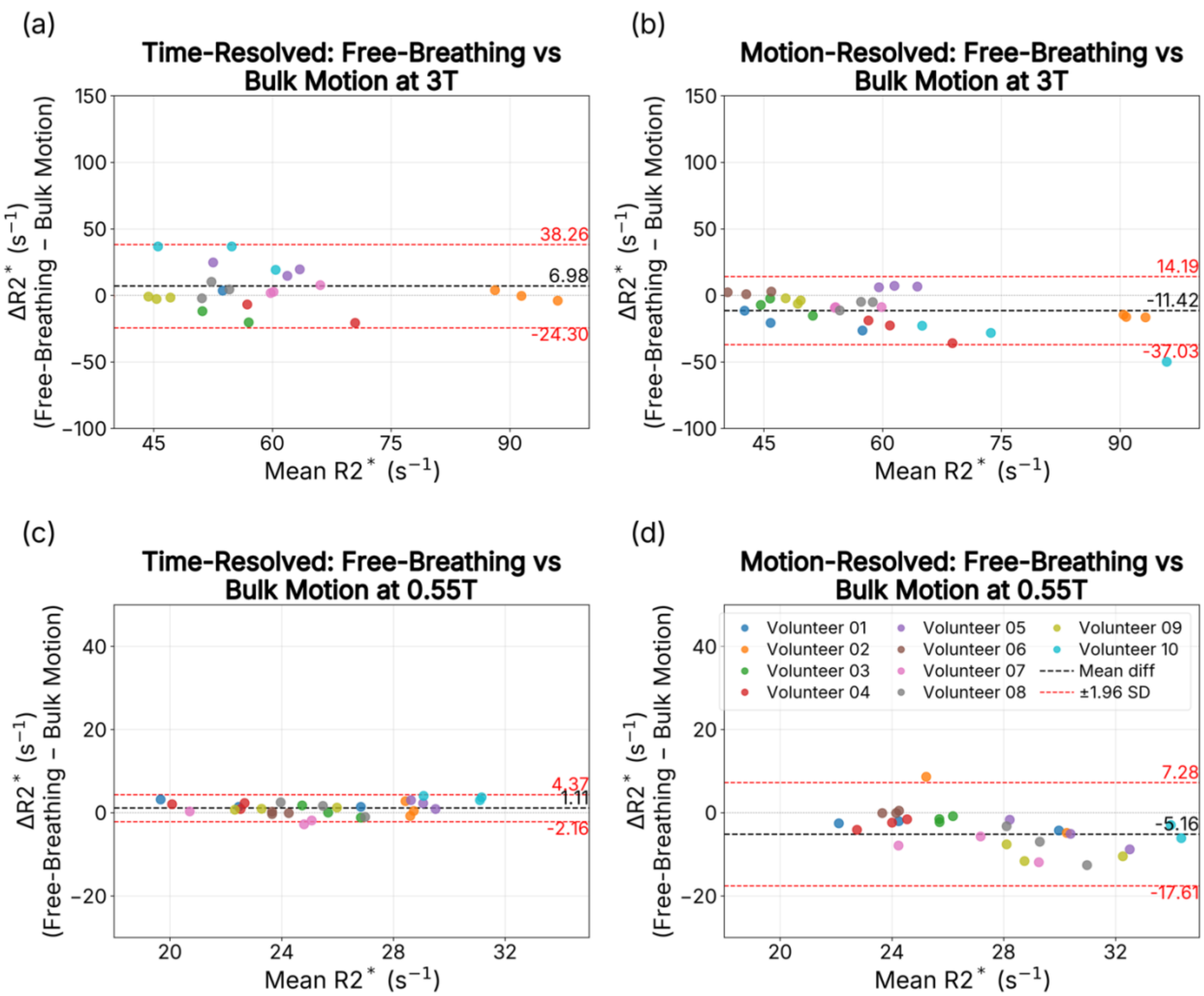

**Figure 10:** The top row compares R2* measurements between normal free-breathing and deliberate bulk-motion acquisitions using time-resolved and motion-resolved reconstructions at 3T, and the bottom row shows the corresponding comparisons at 0.55T. Time-resolved reconstruction showed close agreement between the two motion conditions at 0.55T, with larger differences observed at 3T. Motion-resolved reconstruction showed greater differences between free-breathing and bulk-motion acquisitions at both field strengths.

## Supporting Information Figures

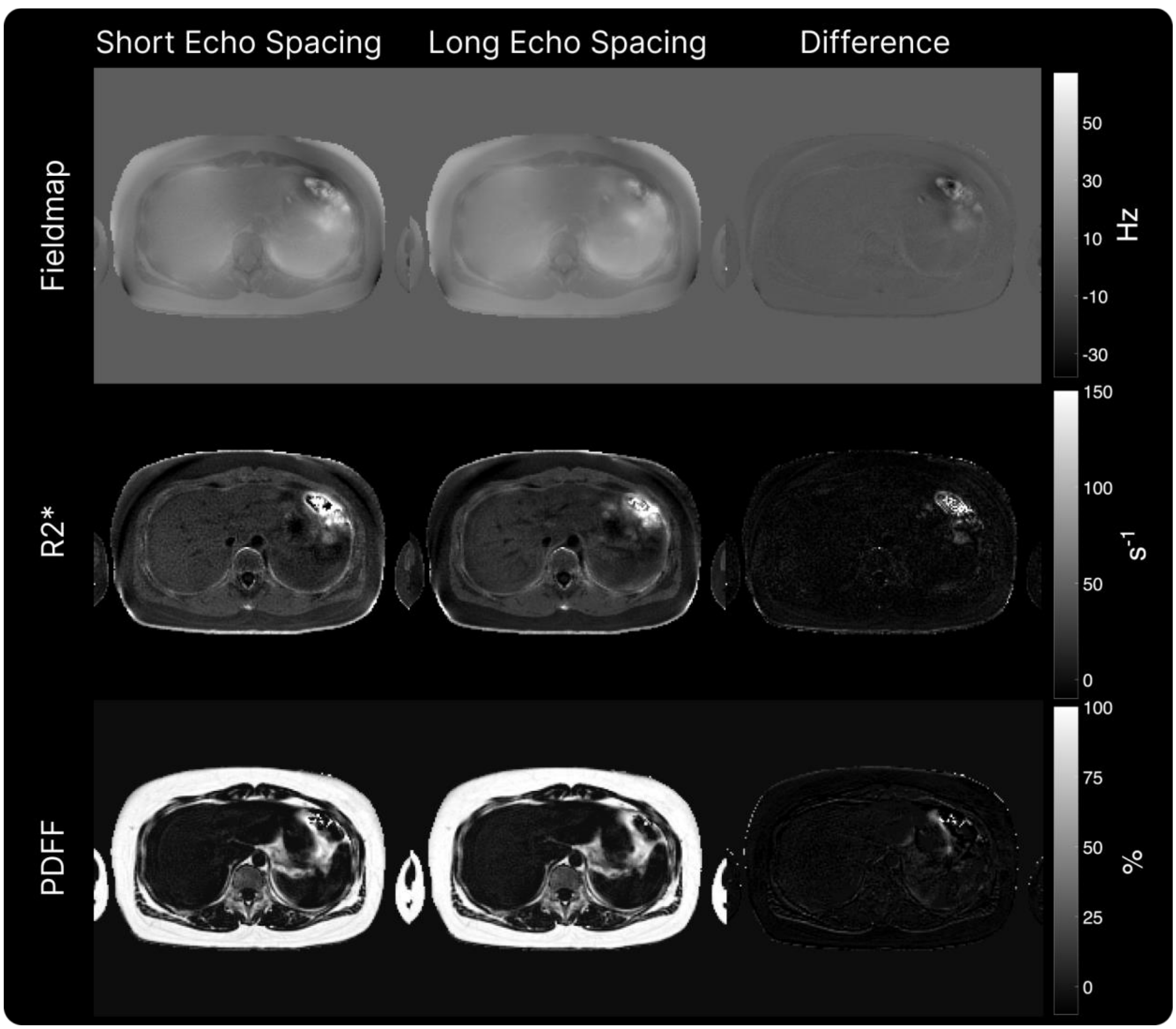

**Supporting Information Figure S1:** B0 field, R2*, and PDFF maps obtained with the shorter and longer echo-spacing settings at 0.55T are shown in a representative subject. The longer echo-spacing setting provided visually less noisy B0 field and R2* maps, whereas PDFF maps were similar between the two settings.

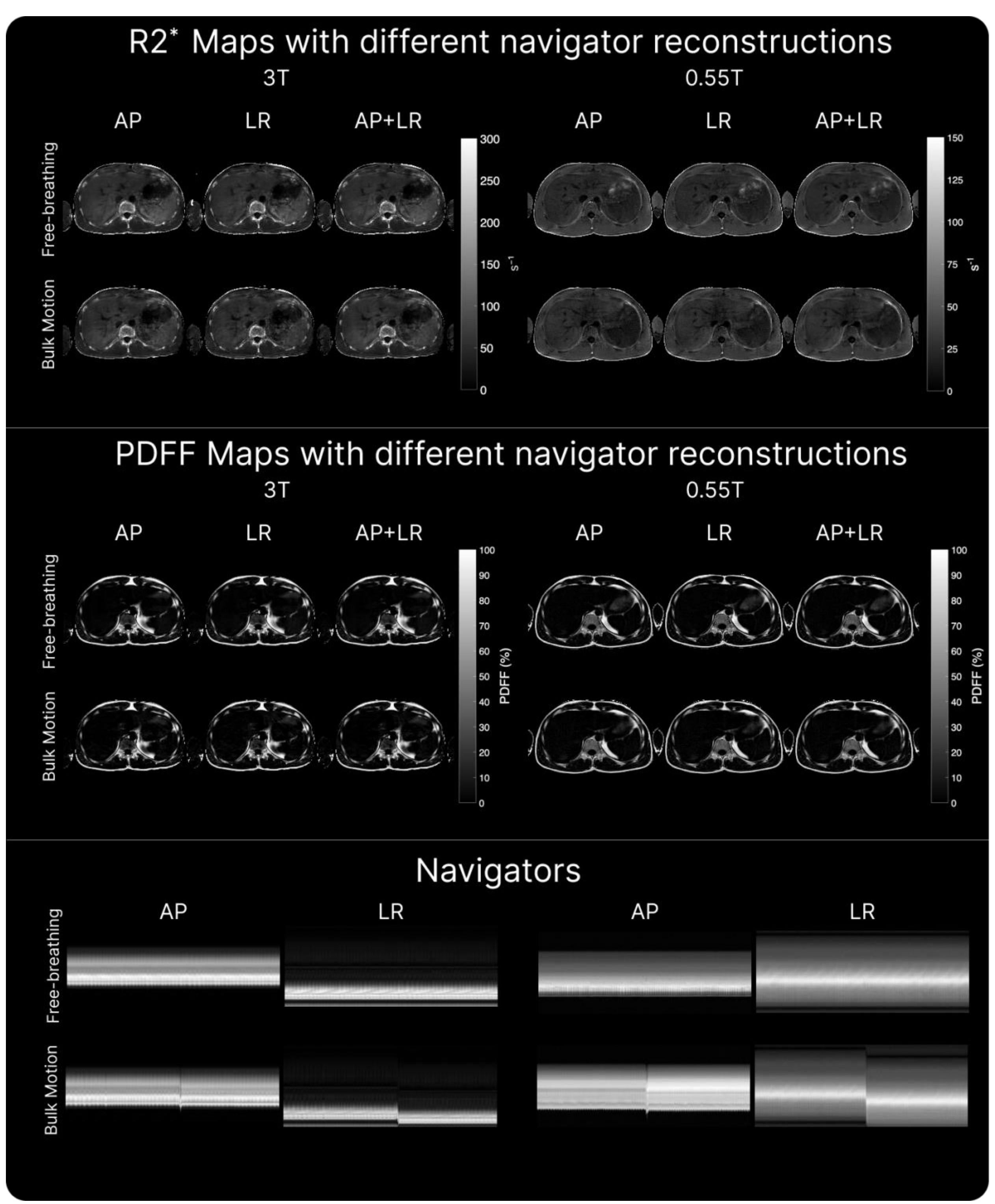

**Supporting Information Figure S2:** R2* and PDFF maps reconstructed using the AP navigator alone, LR navigator alone, and concatenated AP+LR navigators are shown at 3T and 0.55T under normal free-breathing and deliberate bulk-motion conditions. The corresponding AP and LR navigator signals are shown for each field strength and motion condition. For R2*, the combined AP+LR navigator approach provides visually improved image quality compared with either navigator alone, with the differences being more apparent at 0.55T and under bulk-motion conditions. In contrast, PDFF maps show substantially smaller differences among the three navigator approaches at both field strengths and under both motion conditions.

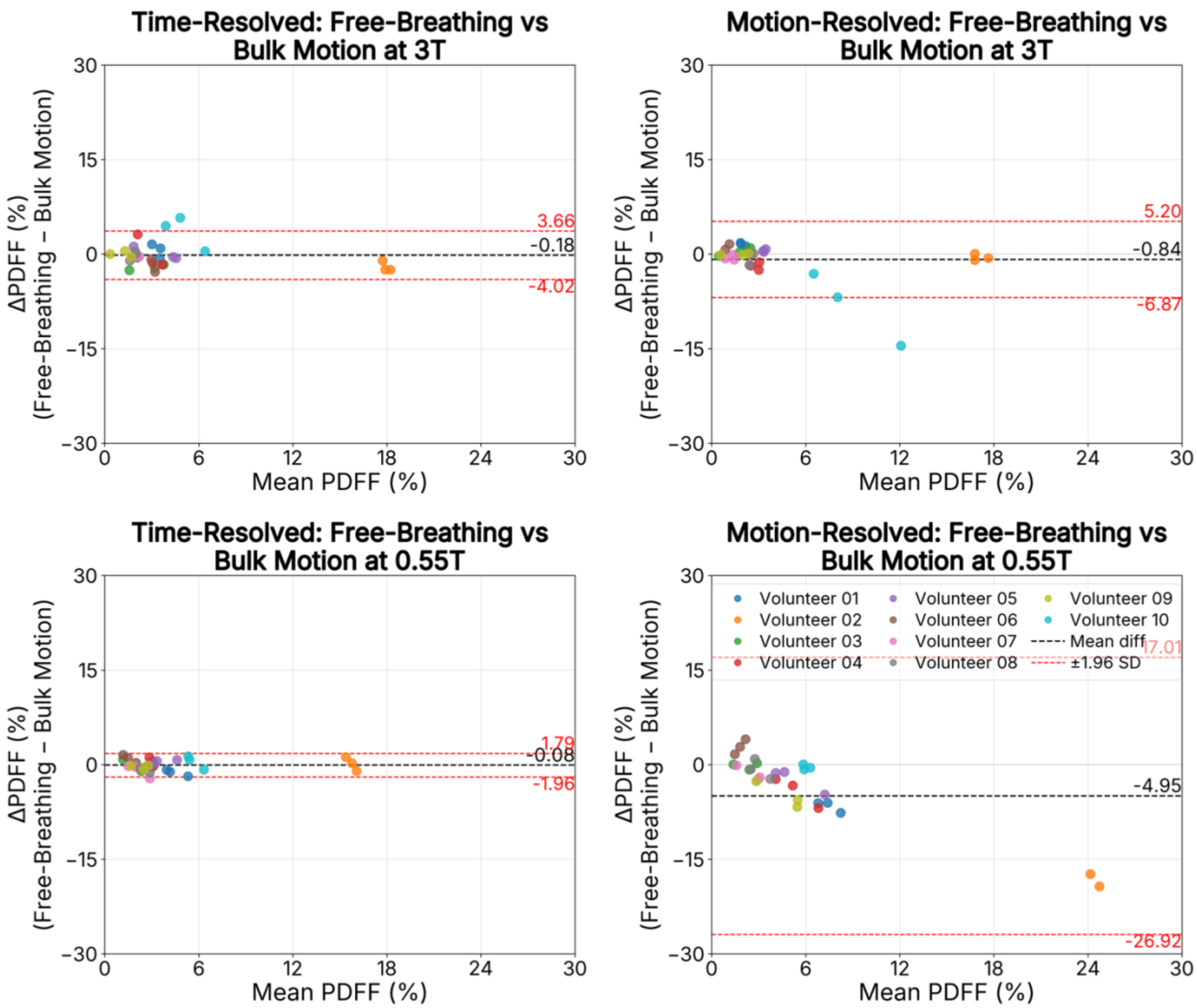

**Supporting Information Figure S3:** The top row compares PDFF measurements between normal free-breathing and deliberate bulk-motion acquisitions using time-resolved and motion-resolved reconstructions at 3T, and the bottom row shows the corresponding comparisons at 0.55T. Time-resolved reconstruction showed close agreement between the two motion conditions at both field strengths, with small biases and narrow limits of agreement. Motion-resolved reconstruction showed larger differences between the two acquisitions, particularly in subjects with higher PDFF at 0.55T.